\documentclass[prl,twocolumn,showkeys,floatfix]{revtex4}
\usepackage{amsmath}
\usepackage{epsfig,amssymb,txfonts}

\newcommand*{\abinit}[0]{\textit{ab~initio}}

\newcommand*{\vasp}[0]{\textsc{vasp}}

\newcommand*{\eqn}[1]{Eqn.~\protect\eqref{eqn:#1}}
\newcommand*{\fig}[1]{Figure~\protect\ref{fig:#1}}
\newcommand*{\tab}[1]{Table~\protect\ref{tab:#1}}
\newcommand*{\pcite}[1]{\protect\cite{#1}}
\newcommand*{\at}[0]{at.\%}
\newcommand*{\TiNbB}[0]{(Ti$_{1-x}$Nb$_x$)B}
\newcommand*{\ordTiNbB}[1]{%
\def\tempa{#1}%
\def\tempb{1}%
\def\tempc{2}%
\def\tempd{3}%
\def\tempe{}%
\ifx\tempa\tempb
(Ti$_{0.75}$Nb$_{0.25}$)%
\fi%
\ifx\tempa\tempc
(Ti$_{0.5}$Nb$_{0.5}$)%
\fi%
\ifx\tempa\tempd
(Ti$_{0.25}$Nb$_{0.75}$)%
\fi%
\ifx\tempa\tempe
(TiNb)%
\fi%
B}

\newcommand*{\x}[0]{\times}
\newcommand*{\EE}[1]{\times 10^{{#1}}}
\newcommand*{\inv}[1]{{#1}^{-1}}

\newcommand*{\conc}[1]{c_\text{#1}}
\newcommand*{\dy}[2]{dy(\text{#1})/d\conc{#2}}
\newcommand*{\dyfrac}[2]{\frac{dy(\text{#1})}{d\conc{#2}}}
\newcommand*{\misy}[1]{\inv{#1}d{#1}/d\conc{X}}

\newcommand*{\delfrac}[1]{\frac{\delta^2}{#1 - \delta^2}}

\newcommand*{\TITLE}[1]{\paragraph*{#1.}}
\newcommand*{\be}[0]{\begin{equation}}
\newcommand*{\ee}[0]{\end{equation}}
\newcommand*{\beu}[0]{\begin{equation*}}
\newcommand*{\eeu}[0]{\end{equation*}}
\newcommand*{\ba}[0]{\begin{array}}
\newcommand*{\ea}[0]{\end{array}}
\newcommand*{\bfig}[0]{\begin{figure}}
\newcommand*{\efig}[0]{\end{figure}}
\newcommand*{\bfigwide}[0]{\begin{figure*}}
\newcommand*{\efigwide}[0]{\end{figure*}}

\newlength{\wholefigwidth}
\newlength{\smallfigwidth}
\newlength{\halfsmallfigwidth}
\newcommand*{\minititle}[2]{\multicolumn{#1}{@{}l@{}}{\hrulefill\ #2 \hrulefill}}
\newcommand*{\entry}[1]{\underbar{\raisebox{1pt}[\depth][\width]{#1}}}

\begin{document}

\title{Lattice and elastic constants of titanium-niobium monoborides
containing aluminum and vanadium}

\author{D. R. Trinkle}
\altaffiliation{Department of Material Science and Engineering,
University of Illinois, Urbana-Champaign, 1304 W. Green Street,
Urbana, IL 61801, USA}
\email{dtrinkle@uiuc.edu}
\affiliation{Materials and Manufacturing Directorate, Air Force Research Laboratory, Wright Patterson Air Force Base, Dayton, Ohio 45433-7817}

\date{\today}

\begin{abstract}
First-principles electronic-structure computes the lattice and elastic
constants of single-crystal TiB and NbB and changes with Nb, Ti, Al,
and V solutes.  The data is built into an interpolation formula for
lattice and elastic constants of the quartenary (\mbox{TiNbAlV})B with
dilute Al and V concentrations.  The lattice and elastic constants of
borides in two Ti alloys containing Nb and Al are predicted from
microprobe measurements.
\end{abstract}

\keywords{Elastic behavior; Titanium alloys; Transition metals;
Borides; Density functional}

\maketitle

\TITLE{Introduction}

Strengthening titanium alloys with titanium-monoboride (TiB) combines
the beneficial stiffness properties of TiB with the plastic behavior
of a titanium alloy matrix, increasing the range of applicability for
automotive and aerospace applications.  The Ti-TiB composites have
increased stiffness, high-temperature strength, creep performance,
fatigue and wear resistance\cite{Godfrey2000,Banerjee2002,Tamir2005}.
Borides also influence the morphology of $\alpha$ phase in
$\alpha$/$\beta$ Ti alloys, producing microstructures similar to
recrystallized alloys\cite{Hill2005}.  Despite the importance of TiB,
accurate predictions of titanium-monoboride elastic constants have
only happened recently\cite{Panda2006}.  Titanium-boride has the FeB
structure: a primitive orthorhombic structure with space group
$Pnma$\cite{Wyckoff}.  The lattice constants measured with x-ray
powder diffraction are $a_0 = 6.12\pm0.01\text{\AA}$, $b_0 =
3.06\pm0.01\text{\AA}$, and $c_0 = 4.56\pm0.01\text{\AA}$
\cite{Decker1954}. The eight-atom unit cell has four equivalent Ti
atoms at the $c$ Wyckoff positions $(0.177,0.25,0.123)$ and four
equivalent B atoms at the $c$ Wyckoff position $(0.029,0.25,0.603)$.
Direct measurements of the 9 independent single-crystal elastic
constants are difficult, and instead elastic-constant combinations
have previously been inferred from Ti-TiB composite
measurements\cite{Atri1999,Gorsse2003}. Panda and Chandran have filled
the gap of experimental measurements with density-functional
calculations of the elastic constants of pure TiB\cite{Panda2006}.

In order to make accurate predictions about the lattice and elastic
constants of monoboride inclusions in titanium alloys, it is necessary
to include the effects of off-stochiometric chemistries sampled in the
monoborides.  These titanium alloys often include aluminum, vanadium,
and niobium additions.  While the borides formed in these alloys are
expected to remain near 50~\at~B composition, potentially any of the
four metals may occupy the metallic sublattice.  Microprobe
measurements of borides in titanium alloys containing niobium have
found a nearly equal concentration of Ti and Nb in the
borides\cite{Cowen2006}.  Metallographic investigation of alloys in the
Ti-V-B system has found comparable V concentration in the metal matrix
and borides after partitioning\cite{Artyukh2006}.  Given the difficulty
of elastic property measurements for the small dispersed borides,
density-functional theory can be used to predict the missing data.
Here, a predictive model of lattice and elastic constants for
(\mbox{TiNbAlV})B in the dilute Al and V limit is constructed using
density-functional theory calculations of TiB and NbB properties,
combined with misfits due to Nb, Ti, Al, and V solutes, all at zero
temperature.  The predictions of the model are compared with a
quasirandom \ordTiNbB{}\ alloys to demonstrate the accuracy of the
interpolation.  Finally, the lattice and elastic constants are
predicted for borides using compositions taken from microprobe
measurements in actual alloys\cite{Cowen2006}.

\TITLE{Methods}

The \abinit\ calculations are performed with
\vasp~\cite{Kresse93,Kresse96b}, a density-functional code using a
plane-wave basis and the projector augmented-wave (PAW)
method\cite{Blochl1994}, with potentials generated by
Kresse\cite{Kresse1999}.  The generalized-gradient approximation of
Perdew and Wang is employed\cite{Perdew91}.  In order to ensure
accurate treatment of the boron potential, a plane-wave kinetic-energy
cutoff of 319~eV is used.  The $k$-point meshes for the 8-atom pure
monoboride cells are $7\x14\x10$ and $7\x5\x5$ for the 48-atom
$1\x3\x2$ supercells, with a Methfessel-Paxton smearing of 0.2~eV.
The PAW potentials for boron and aluminum treat the $s$ and $p$ states
as valence, while the titanium, vanadium, and niobium potentials treat
the $s$, $d$, and filled-$p$ states as valence, corresponding to a
core atomic reference configurations of He for B, Ne for Al, Mg for Ti
and V, and Ca for Nb.  In all cases, the internal atomistic forces are
relaxed to less than 5~meV/\AA, and the stresses are relaxed to less
than 0.2~kbar to determine lattice constants.

\begingroup
\begin{table}
\caption{%
  One volumetric and six volume-conserving strain combinations and the
  linearized stress responses used to compute elastic constants in
  monoborides.  The magnitude of each strain is given by a single
  parameter $\delta$ that can be positive or negative.  The linear
  response of stress to strain is determined by a linear combination
  of elastic constants, and these linear combinations are inverted to
  determine the elastic constants from the electronic-structure
  calculations.  The first strain combination is a purely volumetric
  strain; the remaining six combinations conserve volume for all
  $\delta$.  The first four strain combinations produce stresses
  dependent on $C_{11}$, $C_{22}$, $C_{33}$, $C_{12}$, $C_{23}$, and
  $C_{13}$; the final three give $C_{44}$, $C_{55}$, and $C_{66}$.}
\label{tab:strains}

\parbox{\smallfigwidth}{
\begin{ruledtabular}
\begin{tabular}{cccccc}
\entry{$e_1$}&\entry{$e_2$}&\entry{$e_3$}&
\entry{$\sigma_1/\delta$}&\entry{$\sigma_2/\delta$}&\entry{$\sigma_3/\delta$}\\
$\delta$&$\delta$&$\delta$&
$C_{11}+C_{12}+C_{13}$&$C_{22}+C_{12}+C_{23}$&$C_{33}+C_{23}+C_{13}$\\[3pt]
$\delta$&$-\delta$&$\delfrac{1}$&
$C_{11}-C_{12}$&$-C_{22}+C_{12}$&$-C_{23}+C_{13}$\\[3pt]
$\delfrac{1}$&$\delta$&$-\delta$&
$C_{12}-C_{13}$&$C_{22}-C_{23}$&$-C_{33}+C_{23}$\\[3pt]
$-\delta$&$\delfrac{1}$&$\delta$&
$-C_{11}+C_{13}$&$-C_{12}+C_{23}$&$C_{33}-C_{13}$\\[3pt]
\hline\\[-11pt]
\hline\\[-9pt]
\multicolumn{4}{l}{$e_4=\delta$,\quad $e_1=\delfrac{4}$}&
$\sigma_4/\delta=C_{44}$&\\[3pt]
\multicolumn{4}{l}{$e_5=\delta$,\quad $e_2=\delfrac{4}$}&
$\sigma_5/\delta=C_{55}$&\\[3pt]
\multicolumn{4}{l}{$e_6=\delta$,\quad $e_3=\delfrac{4}$}&
$\sigma_6/\delta=C_{66}$&\\[3pt]
\end{tabular}
\end{ruledtabular}}
\end{table}
\endgroup

The changes in lattice and elastic constants of TiB and NbB with
dilute substitutions of Nb, Ti, Al, and V are extracted from a
systematic series of relaxed and strained supercells, all at 0K.
First, the TiB, NbB, and (48-atom) quasirandom \ordTiNbB{1},
\ordTiNbB{2}, and \ordTiNbB{3}\ lattice constants are determined via
conjugate-gradient minimization.  Special quasirandom
structures\cite{Zunger1990,Wei1990} are periodic supercells where the
chemical species are chosen to most closely approximate a random
structure; c.f., \fig{SQRS}.  The quasirandom structures were
constructed using the program \textsc{atat}~\cite{ATAT}.  To determine
the elastic constants, relaxed cells are subjected to different
magnitudes of one volumetric strain and six volume-conserving strains
(c.f., \tab{strains}); the magnitudes range from $\delta=-0.005$ to
$\delta=+0.005$ in steps of 0.001.  After the atomic degrees of
freedom in each strained cell are fully relaxed, the stresses are
calculated.  Each stress $\sigma_i$ is a linear combination of strains
$e_j$ and elastic constants $C_{ij}$: $\sigma_i = \sum_j C_{ij} e_j$;
c.f., \tab{strains}.  The ratio of stress to $\delta$ from positive and
negative $\delta$ values are averaged.  A range of strains is used to
check that the effect of anharmonicity is negligible; the elastic
constant combinations change by less than 0.1\%\ between
$\delta=0.005$ and $\delta=0.001$.  The values for $C_{44}$, $C_{55}$,
and $C_{66}$ are extracted directly from corresponding strains; the
other six elastic constants $C_{11}$, $C_{22}$, $C_{33}$, $C_{12}$,
$C_{23}$, and $C_{13}$ are extracted from a least-squares fit of the
12 data from the first four strains in \tab{strains}.  To compute
lattice and elastic misfits, $1\x3\x2$ supercells of 24 metal ions and
24 boron ions are constructed; a single Al, V, Ti or Nb ion is
substituted for one metal ion in the cell.  Each 48-atom cell is
relaxed to extract lattice constants; the cells break orthorhombic
symmetry slightly, producing monoclinic strains of less than
$10^{-3}$.  These monoclinic strains are removed before calculation of
elastic constants, which proceeds in an identical manner to the TiB
and NbB cases, using $\delta=\pm0.005$ for the volumetric strains, and
$\delta = \pm0.004$ and $\pm0.002$ for the volume-conserving strains.
The misfit for lattice or elastic constant $y$ of each metal species
X=Al, V, and Nb is numerically extracted as
\be
\frac{1}{y}\dyfrac{TiB}{X} = 
24\cdot\ln\left\{
\frac{y(\text{X}_{1/24}\text{Ti}_{23/24}\text{B}_{24})}{y(\text{TiB})}
\right\}
\label{eqn:misfit}
\ee
for TiB, and similarly for NbB with X=Al, V, and Ti.

The composition of a general monoboride (\mbox{TiNbAlV})B is given in
terms of the atomic compositions of the four metal elements,
$\conc{Ti}$, $\conc{Nb}$, $\conc{Al}$, and $\conc{V}$, under the
assumption that metal ions only occupy the metal sublattice, and boron
ions the boron sublattice.  The alloy concentration is parameterized
in the dilute Al and V limit, assuming a random metal sublattice, with
equal partitioning of Al and V to Ti and Nb sites,
\begin{multline}
\left[\text{Ti}_{(1-x)(1-(\alpha+\beta))}\text{Nb}_{x(1-(\alpha+\beta))}
\text{Al}_\alpha\text{V}_\beta\right]\text{B}:\\
x = \frac{\conc{Nb}}{\conc{Ti}+\conc{Nb}} = 
1 - \frac{\conc{Ti}}{\conc{Ti}+\conc{Nb}};\\
\alpha = \frac{\conc{Al}}{\conc{Ti}+\conc{Nb}+\conc{Al}+\conc{V}};\quad
\beta = \frac{\conc{V}}{\conc{Ti}+\conc{Nb}+\conc{Al}+\conc{V}},
\label{eqn:alloy}
\end{multline}
where $x$ is the relative addition of Nb to Ti, and $\alpha$ and
$\beta$ are the additions of Al and V, respectively.  The dilute Al
and V limit corresponds to $\alpha\ll 1$ and $\beta\ll 1$.  Lattice
and elastic constants for a given monoboride alloy concentration are
interpolated from the limiting TiB and NbB values using the cubic
polynomial in $x$,
\be
\begin{split}
&y(x,\alpha,\beta) =\\
  &\;(1-3x^2+2x^3)\left[y(\text{TiB}) + \alpha\dyfrac{TiB}{Al} +
  \beta\dyfrac{TiB}{V}\right]\\
  &\;+(3x^2-2x^3)\left[y(\text{NbB}) + \alpha\dyfrac{NbB}{Al} +
  \beta\dyfrac{NbB}{V}\right]\\
  &\;+x(1-x)^2\dyfrac{TiB}{Nb} + (1-x)x^2\dyfrac{NbB}{Ti},
\end{split}
\label{eqn:interp}
\ee
where $y$ represents the given lattice or elastic constant, and all
derivatives are evaluated in the dilute concentration limit from 
\eqn{misfit}.  The interpolation formula builds the quartenary
random alloy response from the six dilute binary alloy responses.  It
assumes that the misfits sum linearly, and interpolates properties of
\TiNbB\ as a cubic polynomial in $x$.

\bfigwide
\includegraphics[width=\wholefigwidth]{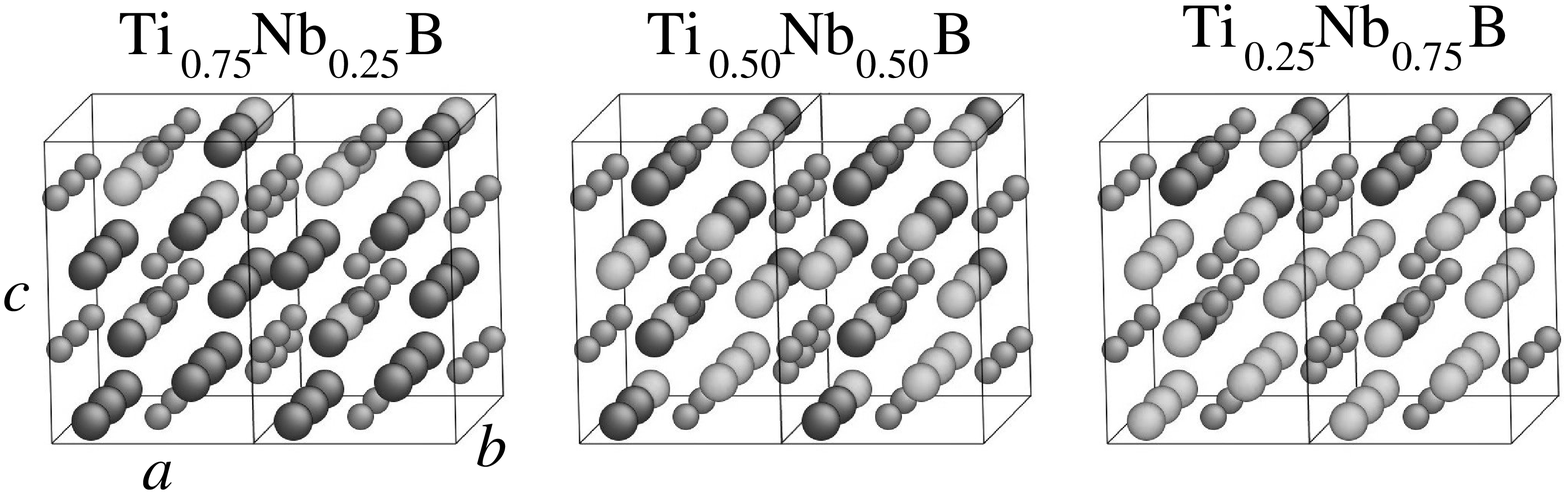}

\caption{%
  Special quasirandom structures for \ordTiNbB{1}, \ordTiNbB{2}, and
  \ordTiNbB{3}.  The 48-atom supercell of Ti (large dark grey), Nb
  (large light grey), and B (small grey) is repeated in the $a$
  direction for visualization.  The chemical identity in each periodic
  supercell is such that the neighbors for each atom approximates a
  truly random structure.  These periodic supercell arrangements allow
  for efficient calculation of properties of a random alloy at the
  given compositions.}
\label{fig:SQRS}
\efigwide

\TITLE{Results}

\bfigwide
\includegraphics[width=\wholefigwidth]{lattice.eps}

\caption{%
  Lattice constants for \TiNbB\ as a function of atomic concentration
  $x$ of Nb.  The values for pure TiB and NbB are shown as the
  endpoints of the interpolation curves.  The solid line shows the
  cubic interpolation of \eqn{interp} for $\alpha=\beta=0$; the
  straight dashed line indicates the result of a purely linear
  interpolation.  The lattice misfits $\inv{y}\dy{TiB}{Nb}$ and
  $\inv{y}\dy{NbB}{Ti}$ are shown italicized with the corresponding
  slopes at the endpoints.  For comparison, the lattice constants of
  the quasirandom \ordTiNbB{}\ alloys are shown as single points for
  $x=$ 0.25, 0.50, and 0.75.  The deviation from the predicted values
  and the quasirandom alloys are due to non-dilute concentrations
  reducing the strength of misfit.}
\label{fig:lattice}
\efigwide

\bfigwide
\includegraphics[width=\wholefigwidth]{elas.eps}

\caption{%
  Elastic constants for \TiNbB\ as a function of atomic concentration
  $x$ of Nb.  The values for pure TiB and NbB are shown as the
  endpoints of the interpolation curves.  The solid line shows the
  cubic interpolation of \eqn{interp} for $\alpha=\beta=0$; the
  straight dashed line indicates the result of a purely linear
  interpolation.  The elastic misfits $\inv{y}\dy{TiB}{Nb}$ and
  $\inv{y}\dy{NbB}{Ti}$ are shown italicized with the corresponding
  slopes at the endpoints.  For comparison, the elastic constants of
  the quasirandom \ordTiNbB{}\ alloys are shown as single points for
  $x=$ 0.25, 0.50, and 0.75.  The deviation from the predicted values
  and the quasirandom alloys are due to non-dilute concentrations
  reducing the strength of misfit.  The general trend is a weakened
  elastic response in the random alloy compared with the linear
  interpolation of the endpoints.}
\label{fig:elas}
\efigwide

\begingroup
\begin{table}
\caption{%
  Lattice and elastic misfits for Al and V substitutions in TiB and
  NbB.  The misfit for a lattice or elastic constant $y$ for solute X
  is defined as $\misy{y}(\conc{X}=0)$.  The values are extracted from
  48 atom monoboride supercells, corresponding to an impurity
  concentration of 4.2~\at\ The overall effect of Al and V additions
  is a volume shrinkage for TiB and NbB.  Al weakens the elastic
  response of TiB and NbB, while V strengthens the elastic response of
  TiB and weakens NbB.}
\label{tab:misfits}

\parbox{\smallfigwidth}{
\begin{ruledtabular}
\begin{tabular}{lcccc}
  &\minititle{2}{TiB}&\minititle{2}{NbB}\\
  &\multicolumn{1}{c}{Al}&\multicolumn{1}{c}{V}
  &\multicolumn{1}{c}{Al}&\multicolumn{1}{c}{V} \\
$\misy{a}$& --0.0004 & --0.0549 &  +0.0502 & --0.0351 \\
$\misy{b}$& --0.0155 & --0.0238 & --0.1305 & --0.0426 \\
$\misy{c}$& --0.0262 & --0.0370 &  +0.0378 & --0.0351 \\
\hline
$\misy{C_{11}}$& --0.5087 &  +0.1292 & --1.0050 & --0.2961 \\
$\misy{C_{22}}$& --0.5067 &  +0.0513 & --0.7241 & --0.2352 \\
$\misy{C_{33}}$& --0.9164 &  +0.2021 & --1.3733 & --0.0992 \\
\hline
$\misy{C_{12}}$& --0.7691 & --0.0390 & --0.1827 &  +0.0932 \\
$\misy{C_{23}}$&  +1.5049 &  +0.5071 &  +0.7557 & --0.0166 \\
$\misy{C_{13}}$&  +0.3567 &  +0.2501 & --0.5439 & --0.6332 \\
\hline
$\misy{C_{44}}$& --0.6064 &  +0.2187 & --0.2334 &  +0.0255 \\
$\misy{C_{55}}$& --1.3375 & --0.0543 & --1.8392 & --0.4602 \\
$\misy{C_{66}}$& --0.9031 &  +0.3123 & --1.2579 & --0.2825
\end{tabular}
\end{ruledtabular}}
\end{table}
\endgroup

\fig{lattice} and \fig{elas} show the lattice and elastic constants
for the random \TiNbB\ alloy as interpolated from the TiB and NbB
endpoints (\eqn{interp}), and compared with quasirandom \ordTiNbB{}\
alloys.  The results for pure TiB ($x=0$) match previous calculations
and experiments\cite{Panda2006,Decker1954}.  The cubic interpolation of
\eqn{interp} captures the effect of misfit for the small $x$ and
$(1-x)$ limits.  There was little difference in lattice and elastic
constants and total energy between the quasirandom structures and the
ordered structures, indicating that ordering will be weak, and has
only a small effect on the properties of the borides.  The quasirandom
\ordTiNbB{2}\ alloy deviates from the prediction from misfits
generally by reducing the effect of misfit, approaching the average of
the TiB and NbB properties.  In general, the cubic interpolation is
not dramatically different from the linear interpolation, with the
notable exception of $C_{22}$.  In this case, the quasirandom alloys
shows significant deviation in both $b$ and $C_{22}$ from the misfit
prediction, especially for \ordTiNbB{2}; these are expected to be
related as decreasing lattice constants are connected to increasing
associated elastic constants.  The large deviation indicates that
experimental measurements of lattice constants and elastic for \TiNbB\
may provide potential insight into the effect of non-dilute
concentrations on lattice and elastic properties.

\tab{misfits} shows the Al and V lattice and elastic misfits in TiB
and NbB.  Generally, Al and V solutes decrease the lattice and elastic
constants of TiB and NbB; the exception is V solutes in TiB, which
have positive elastic misfits.  The reduction of elastic constants
contrasts the effects of Nb and Ti on TiB and NbB, which expand the
lattice (c.f., \fig{lattice}); this suggests that Al and V additions in
the borides offer control of lattice matching by offsetting Ti and Nb
expansion.  The decrease in elastic response of TiB with Al and NbB
with Al and V additions generally matches the Nb and Ti response,
including the positive misfit for $C_{22}$; this suggests less
possible control over elastic response in the borides.  While Al and V
are present in many Ti alloys, the prediction of segregation to the
borides with alloy chemistry and heat treatment is difficult;
inferences from lattice constants are possible.  For example,
measurements of the lattice constants of the borides in Ti-6Al-4V
(10~\at~Al, 4~\at~V) give $a=6.11\text{\AA}$, $b=3.05\text{\AA}$,
$c=4.56\text{\AA}$ \cite{Genc2006}; assuming a sensitivity of $\pm
0.01\text{\AA}$, this predicts a maximum concentration of 8~\at~Al and
3~\at~V in the boride.  Generally, experimental measurement of
monoboride chemistry is crucial data for the prediction of boride
lattice and elastic constants in real alloys.

\begingroup
\begin{table}
\caption{%
  Predicted single-crystal lattice and elastic constants for
  experimentally measured monoboride chemistries\pcite{Cowen2006}.
  Microprobe analysis of monoborides in Ti-15Al-33Nb-5B (\at) and
  Ti-22Al-26Nb-5B (\at) alloys is used to interpolate lattice and
  elastic constants using \eqn{interp}.  As can be seen from
  \fig{elas}, the elastic constants near $x=0.5$ are generally reduced
  below the average response of TiB and NbB.}
\label{tab:microprobe}

\parbox{\smallfigwidth}{
\begin{ruledtabular}
\begin{tabular}{ll@{\quad}ccc}
\minititle{5}{Ti-15Al-33Nb-5B (\at)}\\
\multicolumn{2}{l}{chemistry [\at]}&
\multicolumn{3}{c}{lattice [\AA] and elastic [kbar] constants}\\
Ti& 22.32&
$a= 6.1840$&$b= 3.1262$&$c= 4.6833$\\
Nb& 29.50 &
$C_{11}= 4345$&$C_{22}= 4788$&$C_{33}= 4401$\\
Al& 0.429 &
$C_{12}= 1376$&$C_{23}= 1237$&$C_{13}= 1133$\\
V& --- &
$C_{44}= 1906$&$C_{55}= 1726$&$C_{66}= 2227$\\
B & \textit{balance} &
\multicolumn{3}{l}{\textit{param}: $x= 0.569$; $\alpha= 0.0082$; $\beta=0$}\\
\minititle{5}{Ti-22Al-26Nb-5B (\at)}\\
\multicolumn{2}{l}{chemistry [\at]}&
\multicolumn{3}{c}{lattice [\AA] and elastic [kbar] constants}\\
Ti& 22.23&
$a= 6.1835$&$b= 3.1308$&$c= 4.6853$\\
Nb& 31.46 &
$C_{11}= 4387$&$C_{22}= 4808$&$C_{33}= 4461$\\
Al& 0.003 &
$C_{12}= 1396$&$C_{23}= 1242$&$C_{13}= 1136$\\
V& --- &
$C_{44}= 1907$&$C_{55}= 1756$&$C_{66}= 2249$\\
B & \textit{balance} &
\multicolumn{3}{l}{\textit{param}: $x= 0.586$; $\alpha= 5.6\EE{-5}$; $\beta=0$}
\end{tabular}
\end{ruledtabular}}
\end{table}
\endgroup

\tab{microprobe} contains the predicted lattice and elastic constants
for monoborides in two different Ti alloys, using the monoboride
chemistry from microprobe measurements\cite{Cowen2006}.  The
experimentally measured monoboride chemistry of the alloys both show
nearly equal segregation of Ti and Nb to the borides, and appears to
be nearly independent of the Nb concentration in the alloy.  Al shows
a small segregation to the boride, which validates the treatment of
dilute Al in \eqn{interp}.  In both cases, the final B concentration
is nearly 50~\at; this is expected from the Ti-B phase diagram, which
shows TiB to be a line compound with a homogeneity range of
49--50~\at~\cite{Decker1954,ASM-PD,Ma2004}.  Microprobe measurements do
not provide information regarding the possibility of local short-range
Ti-Ti or Nb-Nb order in the boride.  As \fig{lattice} shows, the
lattice constants are expanded relative to the linear interpolation of
TiB-NbB; \fig{elas} also generally shows a reduction in elastic
constants relative to the linear interpolation of TiB-NbB.
Experimental verification of the predicted lattice and elastic
constants can elucidate the effect of non-dilute concentrations in the
boride lattice, through deviations from the random-alloy predictions.

\TITLE{Conclusion}

The results illustrate the ability of electronic-structure
calculations to predict lattice and elastic constants for real
materials, especially in cases where experimental measurements are
extremely difficult or time-consuming.  The calculations for \TiNbB\
borides and the inclusion of dilute Al and V solutes provides a
database for the prediction of lattice and elastic constants in real
Ti alloys, which in turn can aid in the design of new materials.  To
model increasingly realistic material systems requires modern
computational materials science techniques that investigate the
changes in chemistry produced by solid solution, and the effect on
material properties.

\begin{acknowledgments}
The author thanks D.~Miracle, K.~S.~R.~Chandran, and C.~Woodward for
helpful discussions.  The microprobe data was provided by C.~Boehlert
at Michigan State University.  This research was performed while DRT
held a National Research Council Research Associateship Award at AFRL.
This research was supported in part by a grant of computer time from
the DoD High Performance Computing Modernization Program at ASC/MSRC.
\end{acknowledgments}

\end{document}